# A Design of FPGA Based Small Animal PET Real Time Digital Signal Processing and Correction Logic

Jiaming Lu, Lei Zhao, *Member, IEEE*, Peipei Deng, Bowen Li, Kairen Chen, Shubin Liu, *Member, IEEE*, and Qi An, *Member, IEEE*

*Abstract*—Small animal Positron Emission Tomography (PET) is dedicated to small animal imaging. Animals used in experiments, such as rats and monkeys, are often much smaller than human bodies, which requires higher position and energy precision of the PET imaging system. Besides, Flexibility, high efficiency are also the major demands of a practical PET system. These requires a high-quality analog front-end and a digital signal processing logic with high efficiency and compatibility of multiple data processing modes. The digital signal processing logic of the small animal PET system presented in this paper implements 32-channel signal processing in a single Xilinx Artix-7 family of Field-Programmable Gate Array (FPGA). The logic is designed to support three online modes which are regular package mode, flood map and energy spectrum histogram. Several functions are integrated, including two-dimensional (2D) raw position calculation, crystal identification, events filtering, etc. Besides, a series of online corrections are also integrated, such as photon peak correction to 511 keV and timing offset correction with crystal granularity. A Gigabit Ethernet interface is utilized for data transfer, Look-Up Tables (LUTs) configuration and commands issuing. The pipe-line logic processes the signals at 125 MHz with a rate of 1,000,000 events/s. A series of initial tests are conducted. The results indicate that the digital processing logic achieves the expectations.

## I. Introduction

PET [1][2] is a noninvasive medical imaging technique used to evaluate the metabolism level, biochemical reaction, and functional activity of various organs quantitatively and dynamically. In the field of biomedicine, there is a large amount of researches that is not suitable to be performed directly on the human body. These researches are need to be completed on the animals before extrapolated to the human. Small animal PET [3] [4] is a high sensitivity and resolution PET device for small animal imaging. It is increasingly used in various biomedical researches which are not proper to just conducted on human, such as the development of new therapies and medicine. Animals used in experiments are often much smaller than human bodies, which requires higher spatial resolution of the PET imaging system. We participate the design of a marmoset brain PET system, and are responsible for the Singles Processing Unit (SPU) of the PET system. A diagram of the detectors and the data acquisition system of this PET system is shown in Fig. 1.

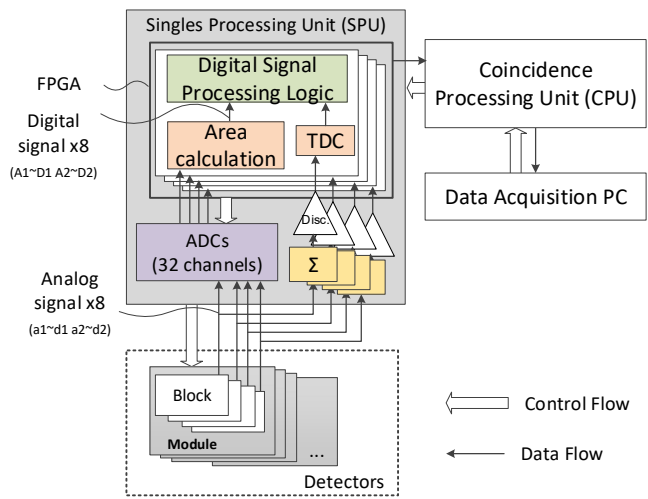

Fig. 1. Diagram of the detectors and the data acquisition system of the small animal PET scanner.

Twelve detector mudules consists a ring which forms the 3D structure of the PET instrument, and each module contains four detector blocks. For one detector block, shown in Fig.2, 23×23 lutetium-yttrium oxyorthosilicate (LYSO) crystal [5][6] array is placed between two layers of Silicon Photomultipliers (SiPM) [7][8]. The eight output signals (a-d1 and a2-d2, four for each end) from one dual-end detector (with which the Depth Of Interaction (DOI) [9][10] can be calculated and leads to a higher spatial resolution in the image reconstruction) are transmitted to SPU. As for the energy and position measurement, the eight signals are transmitted into Analog-to-Digital Convertors (ADCs). Then the eight output digitized signals are transferrd into a Xilinx Artix-7 family of FPGA [11] for area integral calculation. Then the eight ouputs (A1-D1 and A2-D2) of area calculation logic are transferred into the digital signal

Manuscript received Jun. 24, 2018. This work was supported by CAS Center for Excellence in Particle Physics (CCEPP).

The authors are with the State Key Laboratory of Particle Detection and Electronics, University of Science and Technology of China, Hefei, 230026; and Modern Physics Department, University of Science and Technology of China, Hefei, 230026, China (telephone: 086-551-63607746, corresponding author: Lei Zhao, e-mail: zlei@ustc.edu.cn).

processing logic for further processing. As for the time measurement, the eight detector outputs are fed into a analog summation circuit, and the summed signal is discriminated and fed into the FPGA for time measurements. The measurement result of the FPGA Time-to-Digital Convertor (TDC) is then transmitted to the digital signal processing logic. The results of all SPUs are then transferred to a Coincidence Processing Unit (CPU) via Gigabit Ethernet for data coincidence. The CPU finally transfers the data packages to Data Acquisition PC, that communicates with the CPU, SPUs, detector modules and blocks to send commands, configure registers and Look-Up Tables (LUTs) and monitor the status of each part. Fig.3 shows a photograph of an SPU board.

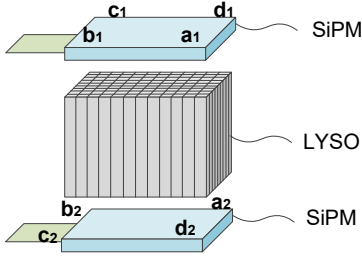

Fig. 2. Diagram of one detector block.

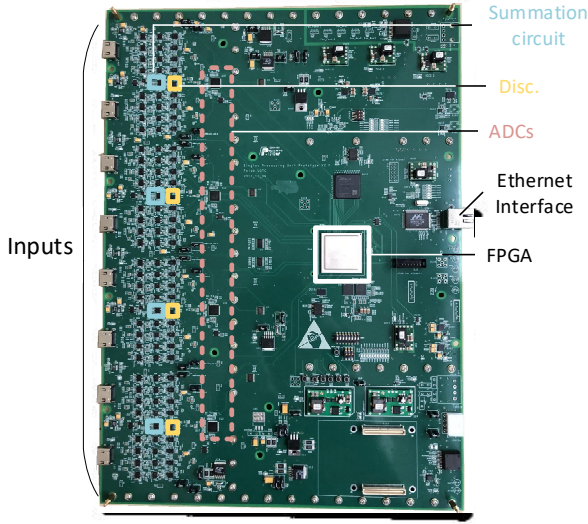

Fig. 3. Photograph of one SPU board.

## II. DESIGN OF SPU DIGITAL SIGNAL PROCESSING LOGIC

High efficiency and flexibility are demanded in a practical PET system. The SPU digital signal processing logic is required to support three modes, which are regular package mode, flood map histogram mode and energy spectrum histogram mode. Regular package mode is mainly used in common imaging process of the small animal PET system. The logic first calculates the raw (x, y) position and the DOI from the eight output signals (A-D1 and A2-D2) of the area calculation logic, and then locates the reaction crystal ID from the raw (x, y) position through a Crystal Look-Up Table (CLT) [12]. On the other hand, to obtain the energy information, the eight input signals are summed to obtain the total energy. Then, photon peak correction to 511keV is performed, and this eliminates the difference of each crystal. For the time information, the output of TDC is corrected to align the difference in the delay of the detector crystals and the corresponding electronic channels (based on crystal ID). Then, the final time information result is output. Flood map histogram mode is mainly used to obtain CLT and energy spectrum histogram mode is mainly used to obtain photon peak LUT (for energy correction).

During the logic design, attention is paid to the following issues: Different functions are required to be switched under the control of PC. Moreover, the online corrections need a sires of LUTs. The two histograms together with the LUTs cost a large amount of memory resource (as shown in Table I). The two on-ine histogram modes and the CLTs of each detector block are the largest ones. Though the usage of the out-chip memory can offer extra memory resources, it may increase the complexity of the design. The solutions of these large logic consumptions within the FPGA are proposed in the following sections of the paper.

TABLE I
MEMORY RESOURCE OCCUPATION OF THE LUTS AND HISTOGRAMS

| LUTs and histograms | Resource occupation (theoretically, 4 block together) / Mb |
|---|---|
| CLTs | 10 |
| Crystal based time offset correction LUTs | 0.07 |
| Photon peak LUTs | 0.06 |
| Flood map histogram | 10 |
| Energy spectrum histogram | 5.16 |
| Total | 25.29 |

As shown in Fig. 3, The SPU digital signal processing logic integrates three modes, as mentioned before. They are switched by the PC by sending command packages through CPU via Gigabit Ethernet. The position, energy and time information, LUTs configuring commands and configuring data are also sent via the same route.

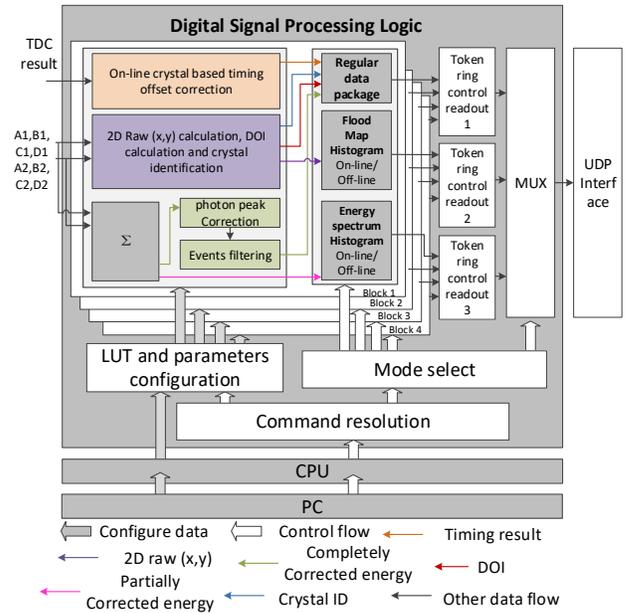

Fig. 3. Diagram of the SPU digital signal processing logic.

### A. Regular Package Mode

Regular signal mode calculates the reaction crystal ID,

corrected charge measurement result (energy) and corrected timing result of each event. It provides the raw information for image reconstruction [13] [14]. The diagram of signal processing of this mode is shown in Fig.4.

For the position information, an 18-bit (9-bit x and 9-bit y) location (raw (x, y)) and a 4-bit DOI are calculated from the 8 outputs of the area calculation circuit by the center-of-gravity method using the following formula (1), (2) and (3):

$$x = 0.5 \times \left( \frac{A1 + D1}{A1 + B1 + C1 + D1} + \frac{A2 + D2}{A2 + B2 + C2 + D2} \right) \quad (1)$$

$$y = 0.5 \times \left( \frac{A1 + B1}{A1 + B1 + C1 + D1} + \frac{C2 + D2}{A2 + B2 + C2 + D2} \right) \quad (2)$$

$$DOI = \frac{A1 + B1 + C1 + D1}{A1 + B1 + C1 + D1 + A2 + B2 + C2 + D2} \quad (3)$$

The raw (x, y) is then used for identifying the interaction crystal with a CLT. The crystal ID is directly output as a part of regular data package.

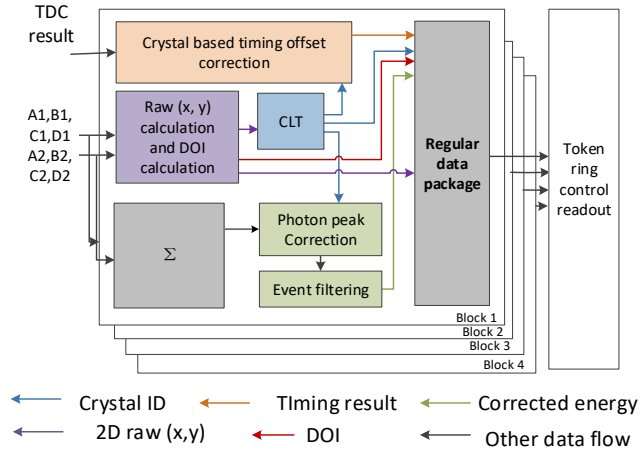

Fig. 4. Diagram of regular package mode.

For the timing information processing, as mentioned before, the crystal based timing offset correction is performed. This adds a specific offset to the measurement result of the TDC according to the reaction crystal ID.

For the energy information processing, the eight signals are summed, and the photon peak correction is performed to eliminate the difference in each crystal. To relieve the data transmission pressure, the events are filtered with a preset energy window to eliminate electronic noise and Compton events before packaging.

Because of the multiple functions of the SPU module that leads to the complexity of output data types, the data type (regular package) information are marked in the regular data packages. In addition, module ID, block ID, etc. are also marked.

The regular data packages of each block are buffered in block FIFOs. There are four detector blocks and the data of each block arrive randomly. If the data of each block is read out at regular intervals, it may happen that the data of a specific block suddenly increases and overflows. To make the data reading proceed properly, a token ring structure is applied.

*B. Food Map Histogram Mode and Energy Spectrum Histogram Mode*

Flood Map is a 2D 512×512 histogram calculated from a large amount of raw (x, y) position data. This is used to generate the CLT. To improve the efficiency of histogram process and achieve online CLT generation in the future design, the online histogram mode, as shown in Fig.5, is integrated. In the on-line mode, the processing of generating the histogram of is conducted in FPGA based on a 10-bit-width 512×512-depth block Random Access Memory (RAM).

The initial state machine starts when the starting command arrives. The current obtained raw (x, y) is stored in the address register, and the event count stored in the corresponding address of the RAM is read out, added with one, and then written back to this address. At the same time, the logic also determines whether the count under this address overflows. If it does, the RAM full flag is set high, and the logic terminates the statistical process; if not, the logic returns to the idle state and waits for the input of next event.

After the statistics is complete, logic reads the data in each address of RAM and sends it out when an instruction is issued from the PC.

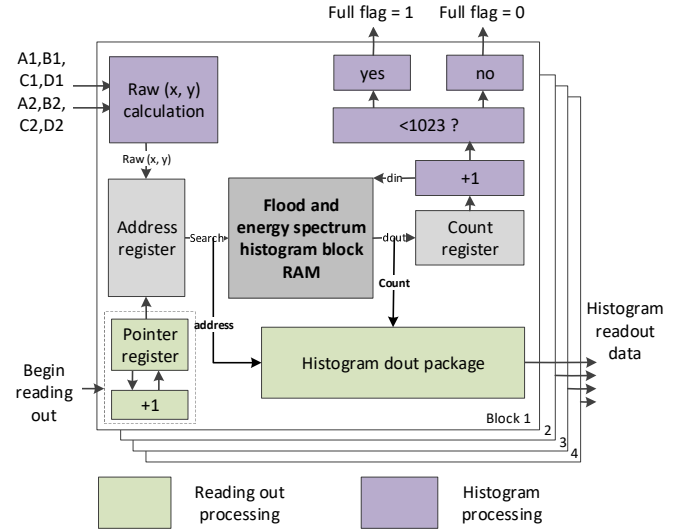

Fig. 5. Diagram of the online flood histogram mode.

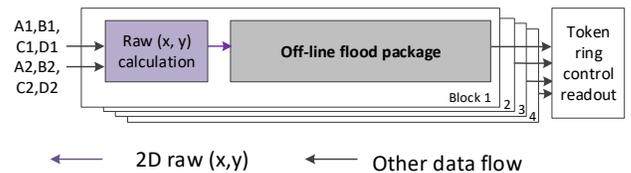

Fig. 6. Diagram of the offline flood histogram mode.

However, for a more flexible and optional system, the backup offline mode is also integrated, as shown in Fig. 6. In this mode,

the raw (x, y) is sent out directly and the processing of generating the histogram is conducted on PC. The online and offline flood histogram mode are able to be switched according to the user's demanding.

Energy spectrum histogram mode calculates the energy spectrum (without the photon peak correction to 511 keV) of each detector crystal to form the photon peak LUT. For the same consideration, both the online and offline sub-modes are integrated and are able to be switched flexibly. The online processing is conducted based on a 10-bit-width 23×23×256-depth block RAM (bigger than the flood) for each detector. The processing of the histogram and readout is similar to the flood.

As the flood and the energy spectrum histogram are not required to be conducted at the same time, we make them share the same 10-bit-width 512×512-depth block RAM source in each block by using data and address multiplexers (MUXs). The used RAM is the size of the flood histogram, as it is the larger one. In this way, a cost of four 10-bit-width 23×23×256-depth block RAMs is reduced, which is the same size as the energy spectrum histogram.

*C. A technical design of CLT*

A typical CLT of one block is shown in Fig.7 (a).

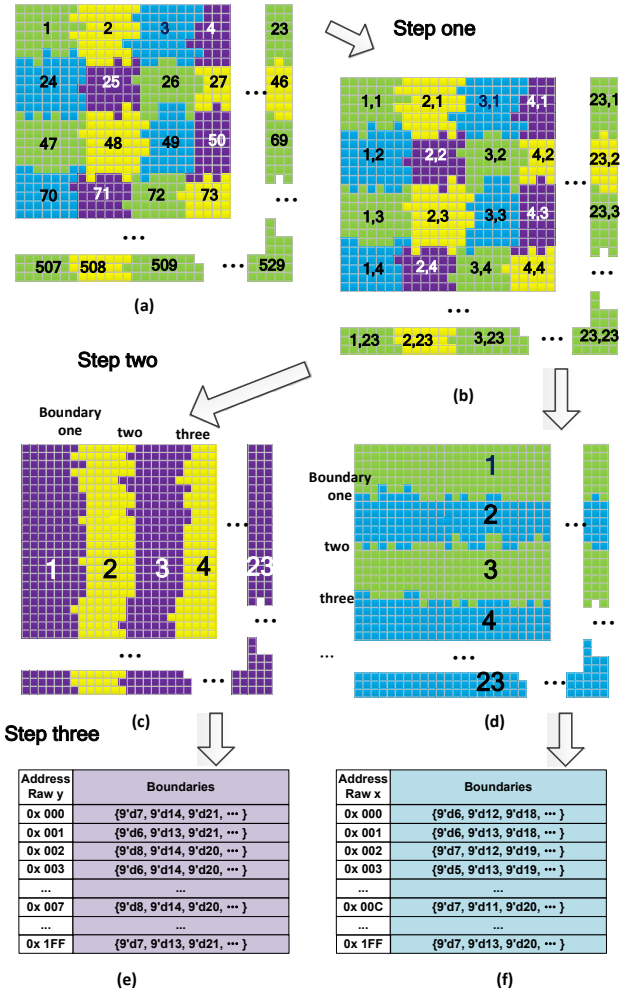

Fig. 7. Transformation from the typical CLT to the boundary CLTs

The smallest squares represent different raw (x, y) positions, and the different areas of different colors represent different crystals. The CLT is stored in a piece of memory resource. Each address (the smallest square in Fig.7 (a)) corresponds to a raw (x, y) position, and the content in this address is the ID of the crystal in which the raw (x, y) position located. Once the raw (x, y) is calculated out, the logic searches this RAM, and then the content (crystal ID) in the corresponding address is directly output.

Given that the binary digits of the raw x or y are n separately, and the amount of crystals is k, a $\lceil \log_2 k \rceil$-width $2^{2n}$-depth memory space ($\lceil a \rceil$ represents the smallest integer larger than the real number a) is requested to store the CLT during the processing. In this small-animal-PET situation, a 10-bit-width 512×512-depth RAM is needed for each block. With such a CLT, a large quantity of repeat information (different raw (x, y) location may corresponds to a same crystal ID) is stored, which, together with other LUTs and the online histograms, causes the RAM resource crisis in FPGA.

A technically design of boundary CLTs is invented to relieve the resource crisis, without using outside-chip memory, which may decrease the integration and increase complexity. Each original CLT is resolved into two boundary CLTs on two directions. The transformation is shown in Fig.7, involving three stages. The first stage transforms the one-dimensional (1D) crystal ID (from Fig. 7 (a)) to 2D (Fig. 7 (b)). The second storage merges the crystals with the same crystal ID component on the two directions (from Fig. 7 (b) to Fig. 7 (c) and Fig. 7 (d)). Several strip combined crystal areas are obtained. In the third stage, the boundary CLTs are obtained. The third-stage process of one direction is shown in Fig.8.

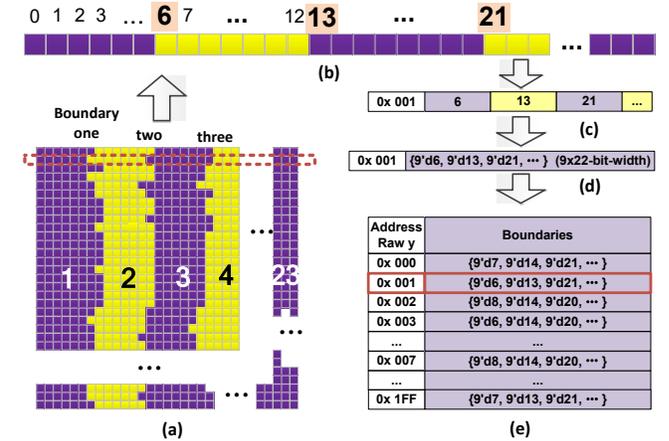

Fig. 8. The third stage of transformation from the typical CLT to boundary CLTs

Each boundary has CLT 22 boundaries. At the boundary, the component of the 2D crystal ID changes (Fig.8 (a)). These boundary information of each direction is stored in a 9×22-bit-width 512-depth RAM (Fig.8 (e)), e. g., each address of the RAM is corresponding to a raw y value, and the boundaries of this line, with the same raw y value, are stored under in address in the form of 22 9-bit-width binary numbers.

The crystal identification with boundary CLTs is conducted as follows (as shown in Fig.9). Raw (x, y) are compared with the boundaries of corresponding direction separately, and the 2D ID components are obtained. Then, a decoder is applied to transform the 2D ID to 1D ID.

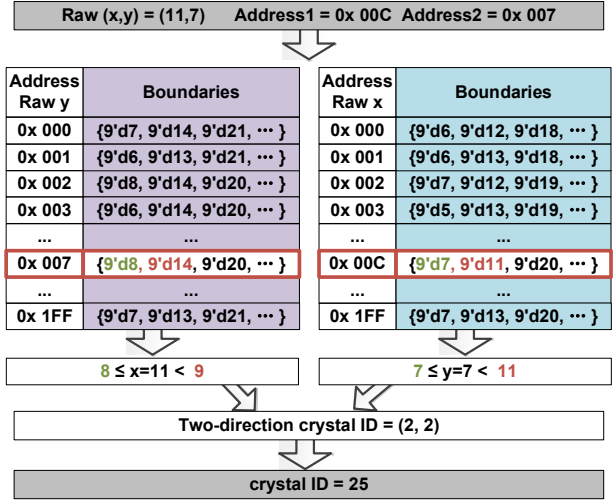

Fig. 9. Crystal identification with boundary CLTs

The storage saving is quite obvious by using the newly designed CLT. The size of new CLT is proportional to $n \times (\sqrt{k} - 1)$ (n is the binary digit of the raw x or y; k is the amount of crystals), whereas the size of the origin one is proportional to $2^n (\lceil \log_2 k \rceil + 1)$, indicating that the newly designed CLT is much more superior when the binary digits of raw (x, y) increase. In this case, the theoretical size of new CLT of one block is 0.19 Mb, whereas the original design is 2.5 Mb, which is 13.16 times as large as the new one.

The memory resource occupations of the optimized and LUTs and the histograms are shown in Table II. The total usage of the memory space is obviously reduced.

TABLE II
MEMORY RESOURCE OCCUPATION OF THE LUTs AND HISTOGRAMS COMPARE

| LUTs and histograms | Resource occupation (theoretically, 4 block together) / Mb | |
|---|---|---|
| | Not optimized | Optimized |
| Boundary CLTs | 10 | 0.76 |
| Crystal based time offset correction LUTs | 0.07 | 0.07 |
| Photon peak LUTs | 0.06 | 0.06 |
| Flood map histogram and Energy spectrum histogram | 10 and 5.16 | 10 |
| Total | 25.29 | 10.89 |

### D. Data Transmission Interface Design

We calculate the event rate of the small animal PET system in order to design the data transmission interface of the SPU.

The probability of gamma photons hitting a detector block is:

$$P_d \approx \frac{25.6^2}{4 \times \pi \times 55^2} = 1.72\% \quad (4)$$

25.6 is the side length of the detector and 55 is the radius of the detector ring.

Assuming a maximum radioactivity of 200 μCi (Curie), i.e., 7.4 MBq (14.8 M singles/s), the detection efficiency of LYSO crystals on gamma photons is 80%. The overall system event rate CR1 and average case rate CR2 for each SPU are:

$$CR_1 = 14.8M \times (1.72\% \times 4 \times 12) \times 80\% \approx 9.78M \quad (5)$$

$$CR_2 = 9.78M / 12 \approx 0.82M \quad (6)$$

The design requires the SPU to have a maximum dead time of 1 μs; therefore, the SPU's maximum tolerance event rate is 1,000,000 singles/s per block, and the total event rate is 4,000,000 bit/s. According to the data requirements, the size of the regular mode data package of each event is 16 bytes. Therefore, the bandwidth requirement under the average event rate is 0.82 M×16 byte = 13.12 MB/s = 105 Mbps, and for maximum event rate, the requirement is 4 M ×16 byte = 64 MB/s = 512 Mbps。

A Gigabit Ethernet interface based on the User Datagram Protocol (UDP) [15] is utilized in data transmission and communication between the CPU and the SPUs. In the design of SPUs, the Transport layer, the Network layer and the Link layer are implemented by the FPGA logic, and the physical layer is implemented by the 88E1111 chip [16] of Marvell Inc. As the data structure between SPUs and CPU is not very complicated, we simplify the standard UDP/IP stack and only necessary functions are implemented.

The diagram of the designed data transmission interface is shown in Fig.10. In a upload (from SPUs to CPU) data transfer process, the data is encapsulated and checked when passing through the Transport layer based on UDP and the Network layer based on Internet Protocol (IP) [17] by the UDP_TX and IP_TX logic and then fed into the MAC_TX logic of the link layer based on Media Access Control (MAC) protocol [18]. The MAC_TX logic adds a part of the MAC header to the data and transfers it to the Tri-Mode-Ethernet-MAC (TEMAC) [19], which is a Xilinx IP core. The IP core performed the most of the functions of the link layer including preamble adding and CRC checking and then transfers the data to the 88E1111 chip through Gigabit Media-Independent Interface (GMII). Then the packages are transferred to the CPU via the RJ45 interface and a cable. In a download transfer processing, when the CPU sends control commands or configuration data to the an SPU, the packages are received and transferred through the same route for parsing, and then the simple commands or data are acquired and transferred to the signal processing logic.

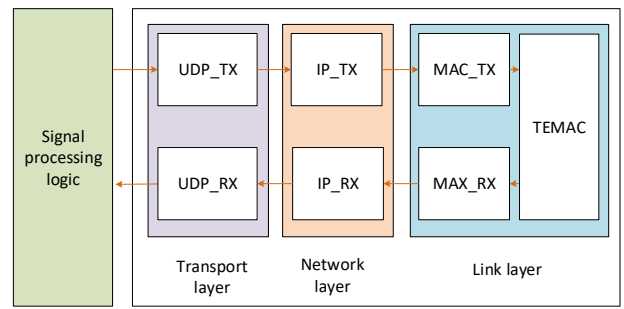

Fig. 10. The crystal identification with the boundary CLTs

## E. Logic occupation

The total FPGA logic occupation is presented in Table III.

TABLE III
LOGIC OCCUPATION

| Resource | Utilization | Available | Utilization % |
|---|---|---|---|
| LUT | 37783 | 133800 | 28.24 |
| LUTRAM | 1573 | 46200 | 3.40 |
| FF | 33236 | 267600 | 12.42 |
| BRAM | 313 | 365 | 85.75 |
| DSP | 5 | 740 | 0.68 |
| IO | 163 | 500 | 32.60 |
| BUFG | 18 | 32 | 56.25 |
| PLL | 2 | 10 | 20.00 |

## III. TESTING RESULTS

### A. Lab Testing

The key performances of SPU module include position precision, time measurement precision, energy (charge) measurement precision and DOI precision. To verify that these performances satisfy the requirements, they are tested in a laboratory environment.

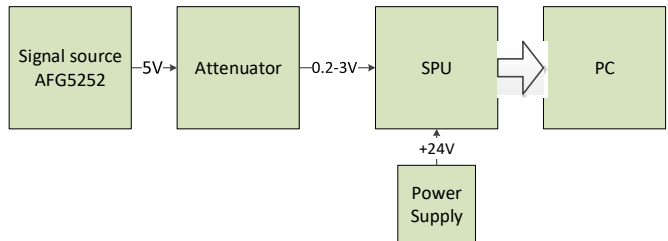

Fig. 11. Diagram of the Lab testing system.

The diagram of the Lab testing system is shown as Fig. 11. The signal source AFG3252 [20] of Tektronix Inc. of output signals of 5.0 V and the signals are transmitted to the attenuator to acquire outputs ranging from 0.2 V to 3.0 V. This is to obtain a better signal-to-noise ratio. Fig.12 shows a photograph of the testing scene.

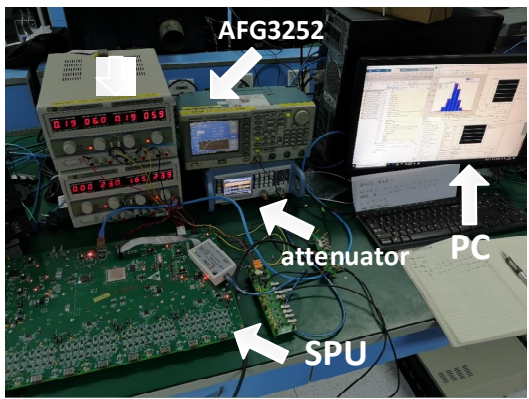

Fig. 12. Photograph of the testing scene.

*1) Position precision*

Technical input signal combinations are generated to simulate different interaction raw positions (center, corners and edges) to conduct the position precision testing.

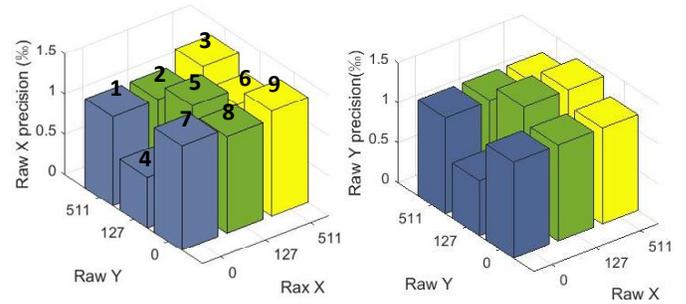

Fig. 13. The position precision of one block (nine positions).

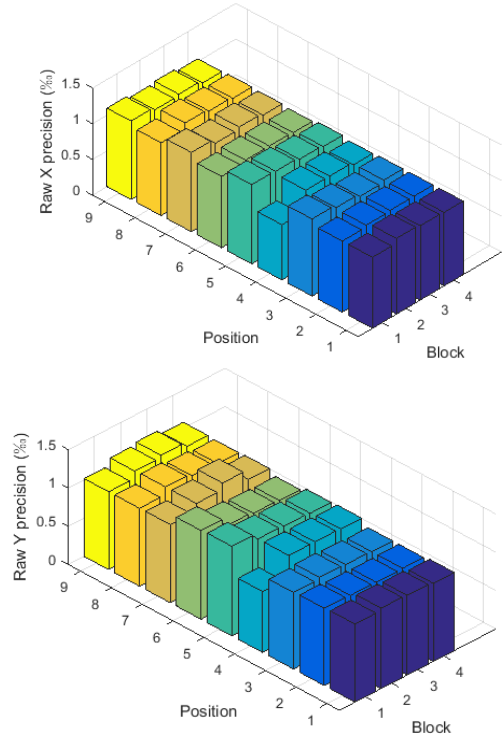

Fig. 14. The position precision of one module.

Fig.13 shows the bar graph of the position testing results of all nine simulated positions of one detector block. Fig.14 shows the bar graph of position testing result of one SPU module including 4 blocks. The testing result shows that the position resolution of each block on each position is better than 1.5‰ RMS.

*2) Timing precision*

To avoid the effect of signal source itself on the time resolution, the delay-line method is applied when conducting the test of time measurement.

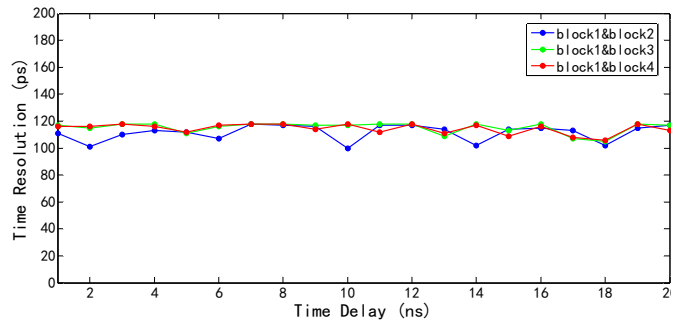

Fig. 15. *Time precision over different time delay.*

An SPU module has four blocks. The time precisions of block 1 & block 2, block 1 & block 3 and block 1 & block 4 are tested separately under different time delay varying from 1 ns to 20 ns. Fig.15 shows the time precision over different time delay of one SPU module, and the time precision under each time delay is better than 118 ps RMS.

*3) Energy (charge) precision*

Fig.16 shows a typical energy testing result of one channel of an SPU module at 2.0 V input, and the energy precision of this channel is 2.4‰ FWHM.

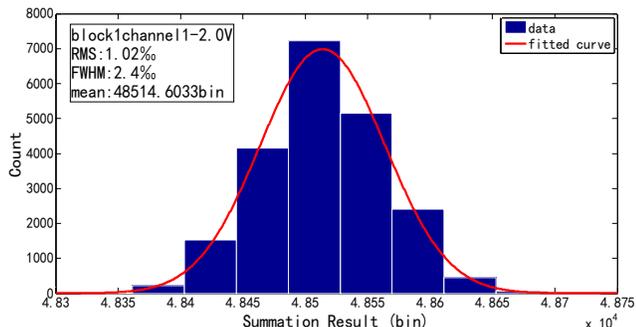
Fig. 16. A typical testing result of energy precision at 2V input.

The energy precision of all 32 channels of one SPU module is tested under the input signal with amplitude ranging from 0.2 V to 3.0 V. Fig.17 shows the results.

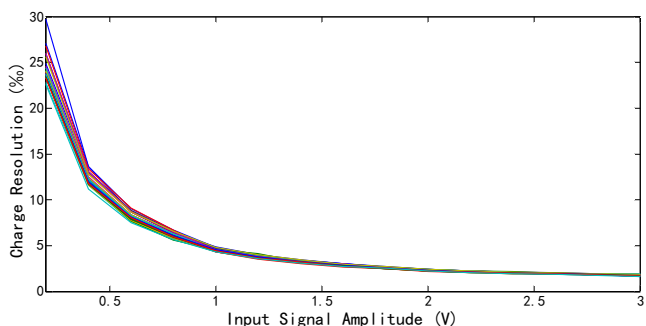
Fig. 17. Energy precision under different input signal amplitude.

*B. Testing with Detectors*

*1) The Flood Map and Energy Spectrum Histogram*

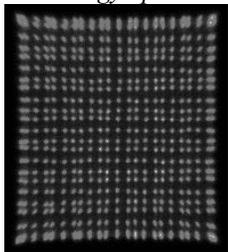
Fig. 18. The 512×512 flood grayscale histogram in online mode.

Fig. 18 shows the 512 × 512 flood grayscale, which is histogrammed in online mode. The grayscale indicates the 23 × 23 LYSO crystals in the block. Fig. 19 shows the online histogrammed energy spectrum of two crystals (ID 76 and ID 79) without photon peak correction to 511 keV. There is difference between the two photon peak positions of the 2 energy spectra.

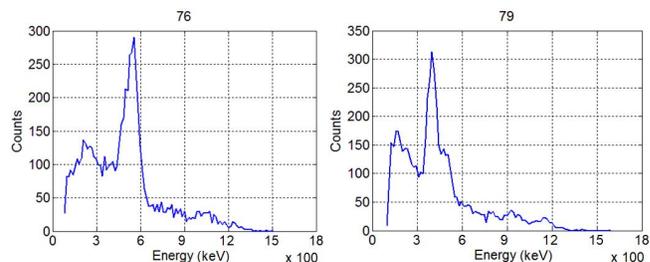
Fig. 19. Online histogram energy spectrum of 2 crystals without photon peak correction.

*2) Photon peak corrrection to 511 keV*

Fig. 20 shows the corrected energy spectra of the two corresponding crystals shown in Fig. 19, which are histogrammed on PC from the regular package mode data. The results indicate that the photon peaks are corrected to 511keV separately.

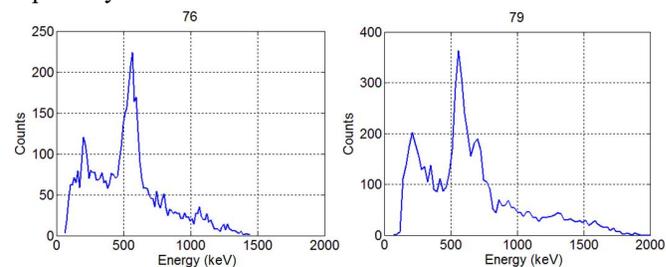
Fig. 20. Offline histogram energy spectrum of two2 crystals with photon peak correction.

## IV. CONCLUSION

The FPGA based small animal PET real time digital signal processing and correction logic is designed. The initial lab testing of SPU module and the initial testing with detectors are conducted. The results indicates that the design achieves the expectations.

ACKNOWLEDGMENT

The authors would like to thank Dr. Yongfeng Yang, Yibao Wu, Xiaohui Wang, Ziru Sang and Zhonghua Kuang in ShenZhen Institutes of Advanced Technology, CAS for their kind help.